\documentclass[aps, prl,nofootinbib,  twocolumn]{revtex4-1}
\pdfoutput=1
\usepackage{epsfig}
\usepackage{amsmath}
\usepackage{amssymb}
\usepackage{subfigure}
\usepackage{bm}   
\usepackage{verbatim} 
\usepackage{natbib}

\newcommand{\be}{\begin{equation}}
\newcommand{\ee}{\end{equation}}
\newcommand{\bea}{\begin{eqnarray}}
\newcommand{\eea}{\end{eqnarray}}

\begin{document}

\title{Bulk-Boundary Duality, Gauge Invariance, and Quantum Error Correction}

\author{Eric Mintun,$^1$ Joseph Polchinski,$^{1,2}$ and Vladimir Rosenhaus$^2$}
\affiliation{$^1$Department of Physics, University of California, Santa Barbara, CA 93106-9530 USA\\ 
$^2$Kavli Institute for Theoretical Physics, University of California, Santa Barbara, CA 93106-4030
USA}

\date{\today}

\begin{abstract} 
Recently, Almheiri, Dong, and Harlow have argued that the localization of bulk information in a boundary dual should be understood in terms of quantum error correction. We show that this structure appears naturally when the gauge invariance of the boundary theory is incorporated.
This provides a new understanding of the non-uniqueness of the bulk fields (precursors).  It suggests a close connection between gauge invariance and the emergence of spacetime.

\end{abstract}

\maketitle


\section{Introduction}

The emergence of bulk spacetimes with gravity from nongravitational boundary theories is a remarkable and fundamental discovery~\cite{Maldacena:1997re}.  A deeper understanding of how local bulk information is encoded in the boundary theory has been the goal of much  work.  It is believed that a subregion of the boundary theory describes a corresponding subregion of the bulk, the causal wedge~\cite{Bousso:2012sj,Czech:2012bh,Hubeny:2012wa}.  Recently, Almheiri, Dong, and Harlow have framed some sharp puzzles in this regard~\cite{Almheiri:2014lwa}.    They note a striking parallel between these and the ideas of quantum erasure correction and quantum secret sharing.


In this paper, we show that quantum error correction and quantum secret sharing appear automatically as a consequence of boundary gauge invariance.  The resulting construction differs from standard examples of error correction in that the precursors act in the full space of gauge-invariant states; it is not necessary to introduce a distinct code subspace.  We illustrate this in two simple models, a discrete model similar to one in Ref.~\cite{Almheiri:2014lwa}, and a free-field model of precursors.   This observation gives a deeper and more general understanding of the nonuniqueness of the precursor construction. While a local bulk operator is dual to many different CFT precursor operators, they are all equivalent when acting on gauge invariant states. 

\section{A discrete model}

The puzzles of Ref.~\cite{Almheiri:2014lwa} deal with ``precursors''~\cite{BDHM,Balasubramanian:1998de,Polchinski:1999yd,Bena99,HKLL}, CFT operators dual to local bulk operators.  The first puzzle is that bulk locality requires a precursor to commute with all spacelike separated local CFT operators.  (To be precise, there are limitations to the commutativity due to gravitational dressing at higher order in $1/N$~\cite{Heemskerk:2012np,Kabat:2012hp,Kabat:2013wga}, which we will discuss later.)  The second puzzle is that one can cover the CFT by three patches $A$, $B$, $C$, such that the precursor can be reconstructed from any of $AB$, $BC$, or $CA$ though not from $A$, $B$, or $C$ separately.  Thus, the bulk information is stored in the CFT in a very nonlocal way.  As Ref.~\cite{Almheiri:2014lwa} notes, this is strongly reminiscent of quantum erasure correction and quantum secret sharing~\cite{QEC}.

\begin{figure}[tbp] 
\centering
	\includegraphics[width=2in]{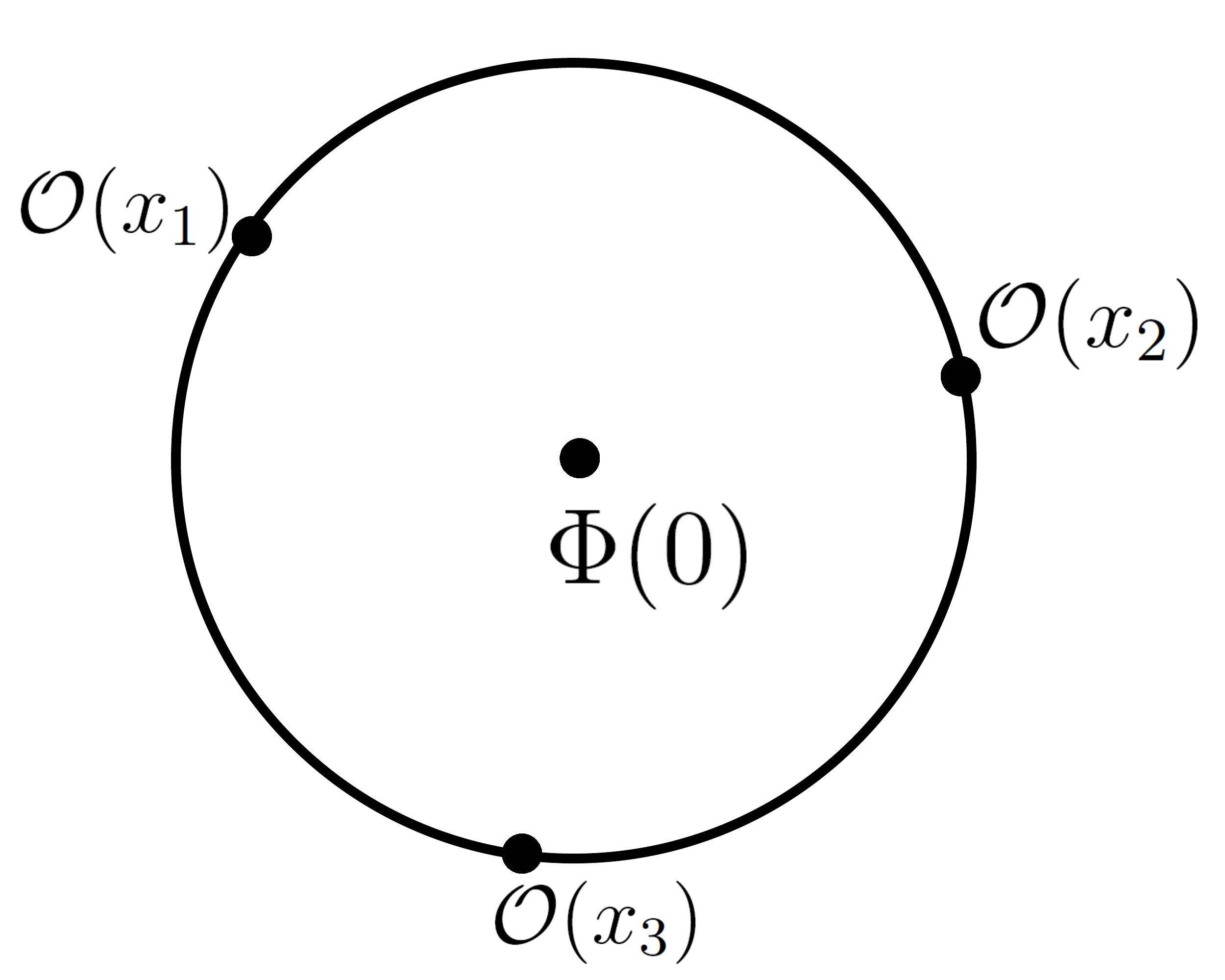}
\caption{Focusing on a single time slice of AdS, bulk microcausality requires the operator $\Phi(0)$ to commute with all spacelike-separated local operators.  By the holographic dictionary, the boundary limits of such bulk operators are local CFT operators $\mathcal{O}(x)$, which must thus commute with $\Phi(0)$.} 
\label{fig1}
\end{figure}

It might appear challenging to construct CFT operators having such properties, but we exhibit here a simple model in which they are realized precisely, and in which gauge invariance plays an essential role.  As in Ref.~\cite{Almheiri:2014lwa}, we model the boundary theory by three sites as in Fig.~1.  The new ingredient is an $O(N)$ gauge symmetry, which for simplicity we take to depend on time but not on space, so that its only effect is to constrain the Hilbert space to invariant states.  In our example, each site has a single bosonic degree of freedom transforming in the $N$ of $O(N)$.  These are denoted $(\phi_a^i, \pi_a^i)$, where the subscript denotes the site and the superscript is the $O(N)$ index;  dot products will always be on the $O(N)$ indices.

We wish to construct a nontrivial precursor operator that commutes with the local gauge invariants, those that are restricted to any single site.  To do this, we note that $\phi \cdot \phi$, $\phi \cdot \pi$, and $\pi \cdot \pi$, which generate the complete set of $O(N)$ invariants on site $a$, form an $SL(2,R)$ algebra.  The doublet $(\phi_a^i, \pi_a^i)$ transforms as the fundamental representation of this algebra, so the antisymmetric combination
\be
L^{ij}_a = \phi^i_a \pi^j_a - \phi^j_a \pi^i_a 
\ee
is a singlet with respect to the $SL(2,R)$ on site $a$, while obviously commuting with the invariants on all other sites.  Thus any linear combination of the operators
\be \label{eq:P}
{\cal P}_{ab} = L^{ij}_a L^{ij}_b 
\ee
is a gauge-invariant operator satisfying the conditions.  These are nontrivial; for example ${\cal P}_{12}$ does not commute with $\phi_2 {\cdot} \phi_3$.  The bulk precursors can then be built out of the ${\cal P}_{ab}$.  For example, ${\cal P}_{12}
+ {\cal P}_{23} + {\cal P}_{31}$ should be $\Phi(0)$.

The ${\cal P}_{ab}$ appear to be double trace operators, while the bulk fields should be single-trace.  However, normal ordering will induce single-trace terms, and these are the leading terms in $1/N$; we will see an example in the next section.  Since the appropriate normal-ordering prescription will depend on the background, a background-independent precursor will have different single-trace parts in different backgrounds.

A better way to understand this construction is to note that  $L^{ij}_a$ is the generator of $O(N)$ transformations acting on site $a$, and so commutes with invariants on any single site.  Any invariant built out of the $L^{ij}_a$ thus commutes with all local invariants.  These are nontrivial in general, as only the total $O(N)$ generator $L^{ij} = \sum_a L_a^{ij}$ annihilates all physical states.
Note also that
\be
{\cal P}_{12} = - {\cal P}_{22} - {\cal P}_{32}  + L^{ij} L_2^{ij} \,.
\ee
It follows that on physical states ${\cal P}_{12}$ is equivalent to $- {\cal P}_{22} - {\cal P}_{32}$, which does not involve the site~1 at all.  

For any operator ${\cal A}^{ij}$ in the adjoint, the combination $L^{ij} {\cal A}^{ij}$ annihilates all gauge-invariant states,
\be
L^{ij} {\cal A}^{ij} \cong 0 \,. \label{gaugefree}
\ee
Note that $L^{ij} {\cal A}^{ij} = {\cal A}^{ij} L^{ij}$.  Also, the commutator of the operator~(\ref{gaugefree}) with an  invariant,
\be
[ L^{ij} {\cal A}^{ij} , {\cal S}] = L^{ij} [ {\cal A}^{ij} , {\cal S}] \,,
\ee
need not vanish identically, but it vanishes on physical states because we can commute $L^{ij}$ through to the right.

The  gauge-invariant precursors thus have natural error-correction built in.  We can read messages encoded in any of the ${\cal P}_{ab}$ by looking at the state on any two sites, while observables on any single site contain no information about any ${\cal P}_{ab}$.  Note that in addition to erasing any single site, we can introduce arbitrary gauge-invariant one-site errors on the remaining sites without losing the message.  This may not fit precisely into the standard framework of quantum error correction~\cite{QEC}. The gauge-invariant subspace in which the precursors act is not a standard code subspace, in that not all states represent distinct messages: it also includes  error degrees of freedom from the one-site gauge invariant operators.\footnote{The code subspace in the model of Ref.~\cite{Almheiri:2014lwa} can also be characterized as a gauge-invariant space, under the $Z_3 \times Z_3$ where the first factor cyclicly permutes the sites and the second shifts all registers equally.  However, there are no analogs of the single-site gauge invariants, which correspond to local operators in the \mbox{CFT}.}

\section{Free-field precursors}

\begin{figure}[tbp] 
\centering
	\includegraphics[width=2in]{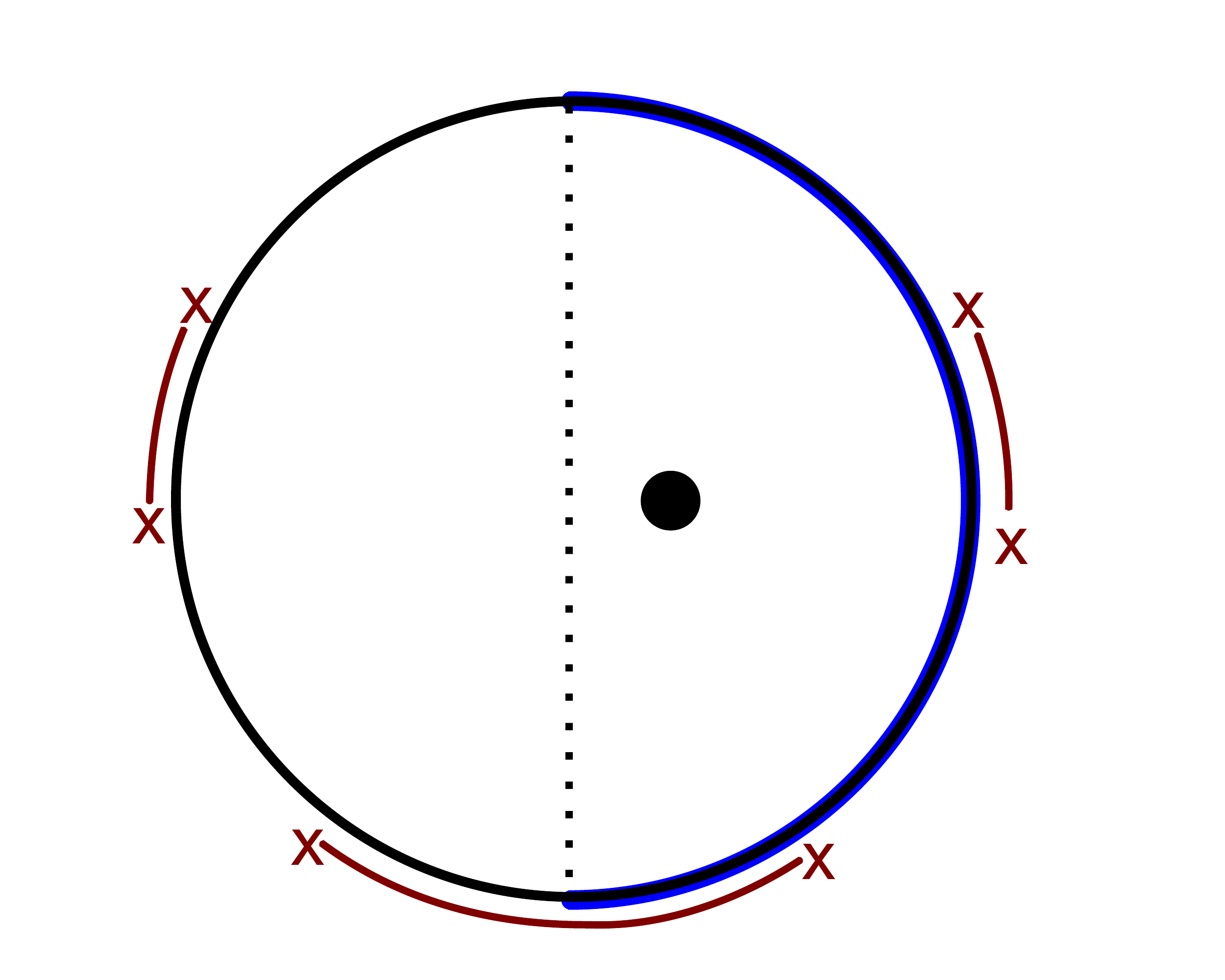}
\caption{A bulk operator (black dot) can be represented as a CFT precursor consisting of bilocals (red arcs ending on x's) stretching over the entire boundary. A different construction represents the bulk operators in the right bulk Rindler wedge as a precursor with support on only the right half of the boundary. Thus, bilocals localized on the left half of the CFT, as well as bilocals stretching from the left to the right half, can be removed from the precursor. This is due to the additional freedom that comes with considering the action of the precursors only on gauge invariant states.}
\label{fig4}
\end{figure}

Now let us apply these lessons to a continuum model of precursors.  Essentially this is an infinite-site version of the preceding example.  We will take the CFT to be a free massless scalar $\phi^i$ in two dimensions; this will not have a fully local holographic dual, but at least at the level of the two-point function it provides a good model \cite{Jevicki,MP}.  The operator $\phi^2$ has dimension zero, so is dual to a massless $AdS_3$ scalar $\Phi(t,\rho,\theta)$.

By the usual precursor construction \cite{BDHM,Balasubramanian:1998de,Polchinski:1999yd,Bena99,HKLL}, one uses bulk equations of motion to express $\Phi(t,\rho,\theta)$ in terms of the CFT field $\phi^2$ smeared over space and time, and then uses CFT equations of motion to evolve this to an operator at a single time. We will work in the large $N$ limit, expanding $\Phi$ in terms of bulk creation operators $b_{\Delta l}^{\dagger}$ for the normal modes $j_{\Delta l}$, 
where $\Delta,l$ are the global energy and angular momentum:
\be \label{eq:pre11}
\Phi(t,\rho,\theta) = \sum_{\Delta l} j_{\Delta l}(\rho, \theta) b_{\Delta l}(t) + \text{h.c.}
\ee
Taking the boundary limit of $\Phi$, we find that $b_{\Delta l}$ is the Fourier transform over space and time of $\phi^2$. Expanding $\phi$ in modes give the precursor
\be \label{eq:balal}
b_{\Delta l} =\frac{1}{\sqrt{N}} \alpha_{\Delta_+} {\cdot} \tilde\alpha_{ \Delta_-} \,,\quad \Delta_{\pm} = \frac12(\Delta \pm l) \,. 
\ee
Here $\alpha_m$  for $m$ positive (negative) is the annihilation (creation) operator for $\phi$ for left-movers of momentum $m$, and similarly for $\tilde{\alpha}_m$ for right-movers (these are normalized $[\alpha_m,\alpha_n] = m \delta_{mn}$), so a bulk particle is dual to a pair of  left and right moving boundary ``particles". The  $1/\sqrt { N}$ normalization is to give the correct $O(1)$ two-point function. Thus (\ref{eq:pre11}) expresses the bulk field in terms of CFT bilinears.
The precursor can alternatively be expressed in position space,
\bea 
\Phi(t,\rho,\theta) &=& \int d\theta'd\theta'' \frac{1}{\sqrt N}\Bigl\{ f(\rho,\theta; \theta', \theta'') \phi(t,\theta') {\cdot} \phi(t,\theta'') \nonumber\\
&& \quad + \ g(\rho,\theta; \theta', \theta'') \pi(t,\theta') {\cdot} \pi(t,\theta'') \Bigr\} \,. \label{prec}
\eea

The integrals~(\ref{prec}) run over the full range $0 \leq \theta',\theta'' \leq 2\pi$, and so determine $\Phi$ in terms of CFT operators on a full Cauchy slice for the \mbox{CFT}.  However, it is expected that bulk physics can be reconstructed in terms of the state on a smaller region of the CFT, namely any one whose causal wedge contains the given bulk point.  One can make this explicit, for example, by separating the CFT into two halves: each  reconstructs a Rindler wedge of the bulk; see Fig.~2.  One can expand the fields in terms of Rindler modes, in a way analogous to (\ref{eq:pre11}). Using now the precursor for a Rindler mode, one has the analogous expression to (\ref{eq:pre11}), but now in terms of the continuous Rindler energy and momenta $\omega, k$ with $\omega_\pm = (\omega \pm k)/2$,
\bea
\Phi(t,\rho,\theta) &=& 
 \int_{ -\infty}^\infty   \frac{d^2\omega_\pm}{\sqrt {N}}h_{\omega_+ \omega_- }^R(\rho,\theta) \alpha_{\omega_+ }(t) {\cdot} \tilde\alpha_{\omega_-}(t)\,. \label{mnprecR}
\eea
The  $\rho,\theta$ are now functions of the AdS-Rindler coordinates, and $h_{\omega_+ \omega_-}^R$ is related to the AdS-Rindler modes.\footnote{There is no AdS-Rindler analog of the position space expression (\ref{prec})~\cite{HKLL}. This is a reflection of the existence of field configurations passing through the bulk wedge and leaving an arbitrarily small imprint on the boundary wedge \cite{BFLRZ}. To reconstruct the bulk on sub-AdS scales, one would need to measure the boundary data to exponential precision \cite{RR}.  While the significance of this on the form of the precursors remains to be understood, in our context the issue is sidestepped by working in momentum space; one could alternatively write a position space expression with complexified boundary coordinates \cite{HKLL} (see also \cite{Morrison:2014jha})}

The Rindler Fourier operators $\alpha_{\omega_\pm}$ (distinguished from the global CFT $\alpha_{\Delta_\pm}$ by their subscript) have support only within the Rindler patch of the CFT, which is half of the CFT.  On the other hand,   the global CFT operators $\alpha_{\Delta_\pm}$, or equivalently the functions~$f(t,\rho,\theta; \theta', \theta'')$ and  $g(t,\rho,\theta; \theta', \theta'')$,  take values in the full Cauchy slice, so that bilinears may lie in either half of the CFT, or stretch between them. So there appears to be a contradiction between the forms~(\ref{eq:pre11}) and~(\ref{mnprecR}).

The point is that there is freedom in the choice of $h_{\Delta_+ \Delta_-}= j_{\Delta l}$.  This is particularly simple in this large-$N$ limit where only the two-point functions of gauge-invariant operators remain. From the form (\ref{eq:balal}), we see that any additional contribution $\alpha_{\Delta_+} \cdot \tilde{\alpha}_{\Delta_-}$ for $\Delta_+\Delta_- <0$, annihilates the vacuum in both directions and so makes no contribution to the two-point function.  (Such an operator would be dual to a bulk mode with $\Delta < l$, which does not exist in global AdS.)  Therefore to this order we have freedom to shift
\be
h_{\Delta_+\Delta_-}(\rho,\theta) \to h_{\Delta_+\Delta_-}(\rho,\theta) + \lambda_{\Delta_+\Delta_-}(\rho,\theta)  \label{hfree}
\ee
for arbitrary $\lambda_{\Delta_+\Delta_-}$ restricted to $\Delta_+\Delta_-<0$.  In the Appendix we show explicitly that the Poincar\'e and Rindler precursors are equal up to the equivalence~(\ref{hfree}).  

This freedom is well-known~\cite{Bena99}, but in light of the preceding section we can express it in a new and powerful way.  The generator of gauge transformations is
\be
L^{ij} = \sum_{r=1}^\infty \frac{2}{r}(\alpha_{-r}^{[i} \alpha_r^{j]} + \tilde\alpha_{-r}^{[i} \tilde\alpha_r^{j]}) +  L_0^{ij}  \,,
\ee
where the last term acts on the zero modes.  Then for $\Delta_+< 0$ and $\Delta_->0$, we have
\be
L^{ij} \alpha^i_{\Delta_+} \tilde\alpha^j_{\Delta_-} = \alpha^i_{\Delta_+} L^{ij} \tilde\alpha^j_{\Delta_-} + (1-N) \alpha_{\Delta_+} {\cdot} \tilde\alpha_{\Delta_-}\,.  \label{normord}
\ee
The first term on the right is normal ordered, and so of order $N$, while the second is of order $N^{3/2}$.  Thus the freedom~(\ref{hfree}) is a special case of the freedom~(\ref{gaugefree}), a surprising result.  What had seemed to be a somewhat obscure dynamical freedom is reinterpreted in terms of gauge invariance. Note too that for $\Delta_+ \Delta_- > 0$ there is no normal-ordering constant in $L^{ij} \alpha^i_{\Delta_+} \alpha^j_{\Delta_-}$, so these bilinears are nonvanishing as we would expect.  This interpretation of the operator identity~(\ref{hfree}) shows that it is not special to the leading large-$N$ behavior, nor to a given vacuum.

In fact, it is not possible for the precursors to commute exactly with all local operators because they cannot commute with the Hamiltonian, which is an integral over $T_{\mu\nu}$.  This corresponds to the need for gravitational dressing to define the bulk position in a coordinate invariant way~\cite{Heemskerk:2012np,Kabat:2012hp,Kabat:2013wga}.  
This did not arise in the free-field model because it is of higher order in $1/N$.  Also at higher orders in $1/N$ there are other complications.  The free-field toy model will not be local in the bulk below the AdS scale.  And, in a realistic model the gauge symmetry will be a function of space as well as time, necessitating Wilson lines in the precursors; their commutators will enter at next order in $1/N$.  In the full-fledged duality, precursors can be constructed order by order in $1/N$ around a semiclassical background~\cite{Kabat:2011rz,Heemskerk:2012mn}, and by construction the localization properties noted above will hold in this expansion.  Extension of the construction beyond this expansion is an interesting question for the future.

The gravitational dressing can be such that the precursor fails to commute with local CFT operators at only one point~\cite{Kabat:2013wga}, for example by working in Fefferman-Graham gauge.  This may seem like too strong a requirement to satisfy, but the precursor construction guarantees it at least in the $1/N$ expansion around a semiclassical background.  It is sometimes argued that this would require the precursors to commute with nonlocal operators as well, because the latter can be expanded in an infinite series of local operators.  However, taking a commutator does not commute with summing such a series.  To see this, consider a free scalar, where
\be
\phi(t,0) = e^{iHt} \phi(0,0) e^{-iHt} = \sum_{n=0}^\infty \frac{t^n}{n!} \partial^n_t \phi(0,0)
\ee
The operator $\pi(0,x)$ commutes with every operator on the right for $x \neq 0$, but only commutes with the operator on the left for $|x|>|t|$.
It should be noted that taking a commutator also does not commute with the operator product expansion, nor with taking the continuum limit of a lattice theory.  The point is that a commutator is the zero-time limit of a difference of time-orderings, and this limit need not commute with the others.

\section{Conclusion}

Dualities only act on gauge-invariant quantities, and so the role of gauge symmetry is often deemphasized.  However, our work suggests a deep connection between gauge invariance in the boundary theory and the emergence of the dual spacetime, which deserves further exploration.   This connection shows up in the natural error-correction that allows for the localization of bulk information in different regions, and which seems to  capture the gauge/gravity structure more precisely than the standard error correction schemes.  This is seen both in the three-site model and in the continuum precursors.  The introduction of gauge symmetries into quantum information theory may be interesting to explore for other reasons as well.

\begin{acknowledgments}
We thank Ahmed Almheiri, Xi Dong and Daniel Harlow for useful discussions, and for comments on the manuscript. This work is supported by NSF Grants PHY11-25915 and PHY13-16748.  

\end{acknowledgments}

\section{Appendix}

Here we show at leading order in $N$ the operator equality of the Rindler and Poincar\'e precursors, up to pure gauge terms.  We use the Poincar\'e description in place of the global description of the main text for ease of notation.

Let $b_{\omega k} = \alpha_{\omega_+} {\cdot} \tilde\alpha_{ \omega_-}/{\sqrt{N}} $ denote the bulk annihilation operator for a Rindler mode with frequency $\omega$ and momentum $k$. 
We do a Bogoliubov transformation to the Minkowski mode operators\footnote{Our notation is such that ${\Gamma}_{\omega\nu}$ includes both the positive and negative frequency Bogoliubov coefficients, depending  on whether $\nu$ is positive or negative.}
\be \label{eq:bBoug}
\alpha_{\omega} = \int \frac{d\nu}{2\pi}\,  {\Gamma}_{\omega\nu}\alpha_{\nu} \,.
\ee
We distinguish the Minkowski modes by their arguments, $\nu,\kappa$.  Then
\be
b_{\omega k} =\frac{1}{\sqrt{N}}  \int\!\!\!\!\int_{-\infty}^\infty \frac{d\nu_+ d\nu_-}{(2\pi)^2}  {\Gamma}_{\omega_+\nu_+}  {\Gamma}_{\omega_-\nu_-}\alpha_{\nu_+} {\cdot}\tilde\alpha_{\nu_-} \,, \label{eq:pre1}
\ee
where $\nu_{\pm} = \frac12(\nu \pm \kappa)$.
Our goal now is to show that this is equivalent to the precursor found from the Poincar\'e smearing function.

The precursor for a Poincar\'e mode is 
\be \label{eq:bpoin}
b_{\nu\kappa} = \frac{1}{\sqrt{N}}\alpha_{\nu_+} {\cdot} \tilde\alpha_{ \nu_-} \,.
\ee
Unlike the Rindler case, where there are evanescent modes with $|\omega|<|k|$ \cite{RR}, the Minkowski modes are restricted to $|\nu|>|\kappa|$, or $\nu_+ \nu_- >0$.
We want to express our AdS Rindler mode in terms of Poincar\'e modes, and then use the precursor (\ref{eq:bpoin}) for each of the Poincar\'e modes. To do this we need to do a bulk Bogoliubov transformation, 
\bea
b_{\omega k} &=&\int\!\!\!\!\int_{-\infty}^\infty \frac{d\nu_+ d\nu_-}{(2\pi)^2}\theta(\nu_+ \nu_-)  \Gamma_{\omega k, \nu\kappa} b_{\nu\kappa} 
\nonumber\\
&=&  \frac{1}{\sqrt{N}}    \int\!\!\!\!\int_{-\infty}^\infty \frac{d\nu_+ d\nu_-}{(2\pi)^2}\theta(\nu_+ \nu_-)  \Gamma_{\omega k, \nu\kappa} \alpha_{\nu_+} {\cdot} \tilde\alpha_{ \nu_-} . \quad \label{eq:pre2}
\eea
In the second line we have inserted the Poincar\'e precursor (\ref{eq:bpoin}).  

The Rindler and Poincar\'e precursors~(\ref{eq:pre1}) and (\ref{eq:pre2}) differ in their ranges of integration, the latter restricted to $\nu_+\nu_- > 0$.  The smaller momentum range translates into a larger coordinate range: the Rindler precursor by construction includes only bilinears with both fields in one Rindler patch of AdS, while for the Poincar\'e precursor each field can be in either patch.  The equivalence of Eqs.~(\ref{eq:pre1}) and (\ref{eq:pre2}) requires that 
\be 
\Gamma_{\omega k, \nu\kappa} =  {\Gamma}_{\omega_+\nu_+}  {\Gamma}_{\omega_-\nu_-}  \label{bogobogo}
\ee
 for $\nu_+ \nu_- > 0$.   The  remaining difference, from modes with $\nu_+ \nu_- < 0$, is then pure gauge to leading order in $1/N$.  This was shown in Eq.~(\ref{normord}) for the global case; the same argument applies for Poincar\'e.

For completeness we verify Eq.~(\ref{bogobogo}). First, we note that near the boundary AdS-Rindler asymptotes to Rindler. To see this, let the AdS-Rindler metric be
\be
ds^2 = - (r^2 - 1) d\tau^2 + \frac{dr^2}{r^2 - 1} + r^2 dx^2
\ee
and consider the coordinate transformation
\be
u = -\frac{\sqrt{r^2-1}}{r} e^{x- \tau}\,,  \ \ v =  \frac{\sqrt{r^2-1}}{r} e^{x+ \tau}\,, \ \ 
z = e^{x}/r \,,
\ee
which yields the Poincar\'e metric, 
\be
ds^2 = \frac{- du dv + dz^2}{z^2} \,.
\ee
Notice that as $r\rightarrow \infty$, we have that 
\be
u\rightarrow - e^{x - \tau}, \ \ \ \ v\rightarrow e^{x +\tau} \,,
\ee
which is  the standard transformation between Minkowski and Rindler coordinates.

The bulk Bogoliubov coefficients are given by the Klein-Gordon inner product between an AdS-Rindler mode and a Poincar\'e mode. This  is usually computed on a spacelike codimension-one surface. We will instead compute the Klein-Gordon inner product  on a timelike codimension-one surface: the boundary of AdS. The boundary limit of the Poincar\'e/AdS-Rindler modes is, respectively,
\be
e^{-i u\nu_+ - i v\nu_-}~,  \ \ \ (-u)^{i \omega_+} v^{i \omega_-}
\ee
where we have rescaled by the conformal factor $z^{\Delta}$ and have chosen a nonstandard normalization for the modes. The Bogoliubov coefficients are thus
\bea
\Gamma_{\omega k, \nu\kappa} &=& \int du dv \, e^{-i u\nu_+ - i v\nu_-} (-u)^{i \omega_+} v^{i \omega_-} \nonumber\\
&=& \Gamma_{\omega_+, \nu_+}\Gamma_{\omega_-, \nu_-} \,,
\end{eqnarray}
 confirming the needed relation (\ref{bogobogo}) between the AdS-Rindler/Poincar\'e Bogoliubov coefficients and the boundary Rindler/Minkowski Bogoliubov coefficients.

\end{document}